\documentclass[12pt]{article}
\usepackage{graphicx}

\def\fun#1#2{\lower3.6pt\vbox{\baselineskip0pt\lineskip.9pt
  \ialign{$\mathsurround=0pt#1\hfil##\hfil$\crcr#2\crcr\sim\crcr}}}

\newcommand{\Li}{\mbox{Li}_2}
\newcommand{\dd}{\mbox{d}}

\newcommand{\vecc}[1]{\mbox{\boldmath $#1$}}

\newcommand{\matr}[1]{\mbox{#1}}
\newcommand{\be}{\begin{equation}}
\newcommand{\ee}{\end{equation}}
\newcommand{\ba}{\begin{eqnarray}}
\newcommand{\ea}{\end{eqnarray}}

\title{ (Quasi)elastic electron--muon large-angle scattering
to a two-loop approximation: vertex contributions
\footnote{supported in part by RFBR 01-02-17437.} }

\author{
V.V.~Bytev, E.A.~Kuraev, B.G.~Shaikhatdenov
\thanks{ on leave of absence from IPT, Almaty-82, Kazakhstan. }
\vspace{4mm}
\\
\small\sl Joint Institute for Nuclear Research, Dubna, 141980, Russia}

\date{\today}
\begin{document}
\maketitle

\begin{abstract}
We consider a process of quasielastic $e\mu$ large-angle scattering at high energies
with radiative corrections up to a two-loop level.
A lowest order radiative correction arising both from one-loop virtual photon emission
and a real soft emission are
presented to a power accuracy. Two-loop level corrections are supposed to be of three
gauge-invariant classes. One of them, so called vertex contribution,
is given in logarithmic approximation. Relation with the renormalization
group approach is discussed.
\end{abstract}

%--------------------------------------------------------------
\section{Introduction}

Certain interest to the physics envisaged at electron-muon colliders is now surging up. The main attention will be
paid to the investigation of rare processes, for instance to those which violate the lepton number conservation law.
Another motivation is a test of the models alternative to the SM~\cite{BPT}. The problems of calibration and precise
determination of luminosity will be important. To this end the process of quasi-elastic electron-muon scattering
could be used.

Processes of quasi-elastic and inelastic large-angle $e\mu$ scattering (EMS)
play an important role in the luminosity calibration
at electron-positron colliders. Indeed they have a clear signature:
scattered leptons moving almost back-to-back (in the center-of-mass (cms) reference frame)
and sufficiently  large cross section ($\theta$- cms scattering angle ),
\ba
\frac{\dd\sigma_0(\theta)}{\dd\Omega_e}\simeq\frac{200~{\matr{nb}}}{s~({\matr{GeV}}^2)},
\qquad cos \theta\approx\frac{1}{2}.
\ea

The modern experimental requirements to the theoretical accuracy
are at the level of per mille or even less whereby demanding for a detailed knowledge
of the nonleading terms in two-loop approximation.
Some of them have been recently calculated in a series of papers~\cite{Bhabha} devoted to
the study of large-angle Bhabha scattering.
The contribution of elastic genuine two-loop virtual correction to the ~Bhabha amplitude has been recently
performed~\cite{GTB} using the prescription developed in~\cite{Cat} how to handle singular terms in QCD at two-loop level.

In this paper we consider the ~EMS process to a two-loop approximation.
We are interested in the contribution to the cross section at this level which is given
by the interference of Born amplitude and those that take into account the two-loop virtual corrections to the former.
An attempt to this problem was done in a series of papers~\cite{FO} where a direct calculation
was performed, but unfortunately their result is incorrect even in the part containing infrared (IR) divergence.
Another set of papers (see for example~\cite{SV}) was devoted to the calculation of two-loop Feynman amplitudes within
a dimensional regularization scheme. Once again their results cannot be straightforwardly applied to the real
amplitudes of $e\mu$ large-angle scattering. One of the reasons is the requirement of distinguishing
different masses of interacting particles.

\begin{figure}[htbp]
\begin{center}
\includegraphics[]{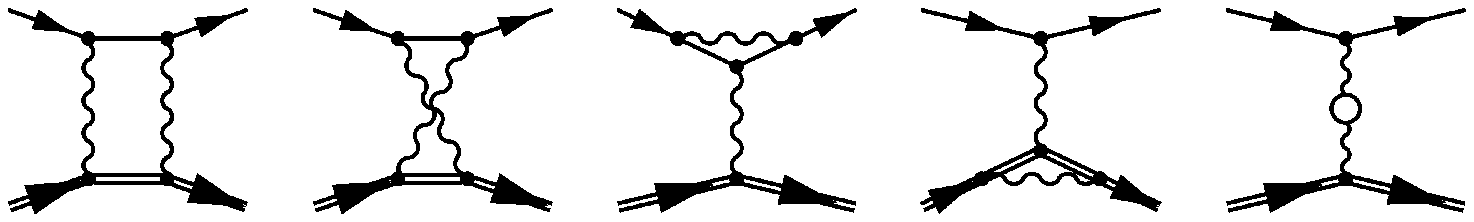}
\end{center}
\caption{ First order contributions.}
\label{fig1}
\end{figure}

Here we will consider only virtual and real soft photon contributions to the cross section of $e\mu$ scattering.
To a third order of PT there exist three sets of contributions, each of which is free of IR singularities.
They include the contribution coming from the one-loop virtual photon emission corrections
(see Fig.~\ref{fig1}) and the one given by a soft photon emission (see ~Fig.\ref{fig3}a).
%While evaluating them we omit effects of fermion identity keeping in mind that the corresponding
%contribution is of order ${\cal O}(1)$.
%--------------- fig ---------------
%--------------- fig ---------------

To a fourth order there in overall are four IR-free sets. One of them, dubbed vertex,
contains virtual corrections up to a second order of PT to the lepton vertex function
and relevant inelastic
processes with emission and absorption of real soft photons and lepton pairs by initial
and scattered electron (and the same for muons). We use here the known expression for
the lepton vertex function up to a fourth order of PT~\cite{BMR}.
These along with the contribution coming from the emission of two real soft photons and
soft charged lepton pair
is our primary concern in the present paper. As well we consider a contribution to the
vacuum polarization caused by hadrons and a soft real pion pair production.

Three additional gauge invariant contributions are described
by the one photon exchange containing lepton vertex functions
with account for the vacuum polarization (VP) and box-type FD with a self-energy insertion
into the one of the exchange photons' Green function. They are put aside for a separate
consideration.

Quasielastic means a process with emission of final particles in a center-of-mass (cms) reference frame
almost back-to-back. Final particle's energies up to a small value $\Delta\varepsilon\ll\varepsilon$ coincide
with those of initial particles. This disbalance is due to a possible emission of soft photons and pairs.

We start by giving the results for the Born differential cross section and first order corrections.
The latter contains radiative corrections (RC) due to emission of virtual photons at one-loop level and
emission of an additional soft photon. Those contributions suffer from IR
divergence that mutually cancel out upon summing them up.
%--------------- fig ---------------
\begin{figure}[!thb]
\begin{center}
\includegraphics[]{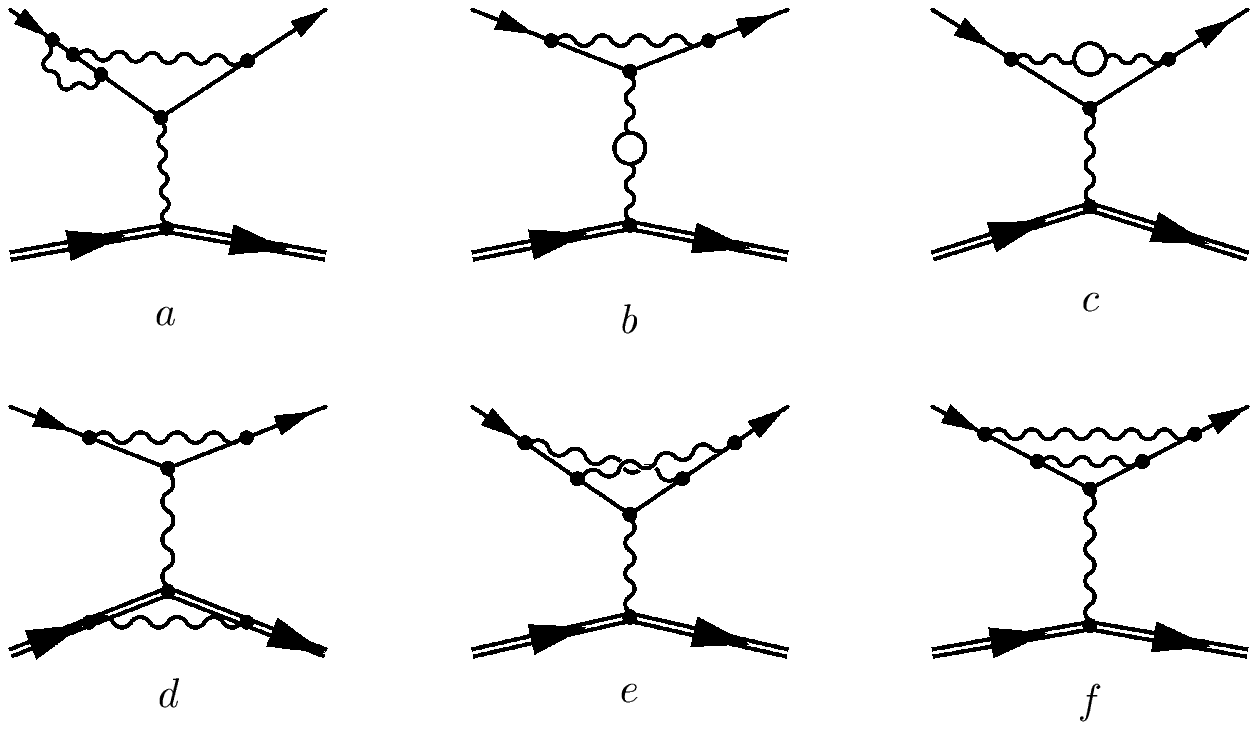}
\end{center}
\caption{ Some of V-type second order contributions. }
\label{fig2}
\end{figure}
%--------------- fig ---------------

An outcome of the calculations agrees with the renormalization group (RG) prediction
in the leading logarithmic approximation (LLA):
\ba\label{eq1}
\frac{\alpha}{\pi}\rho_t\sim 1,\qquad \frac{\alpha}{\pi}\ll 1,
 \qquad \rho_t=\ln\frac{-t}{m_em_\mu},\nonumber \\
\dd\sigma=\frac{\dd\sigma_0}{|1-\Pi(t)|^2} {\cal D}_{\Delta}^4,
\ea
with ${\cal D}_{\Delta}$ is the $\Delta$-part of non-singlet structure function of
lepton \cite{FK}
\ba
\label{eq1a}
&& {\cal D}_{\Delta}=1+\sum\limits_{n=1}^\infty\left(\frac{\alpha}{2\pi}
\rho_t\right)^nP_{n\Delta},
\\ \nonumber
P_{1\Delta}=2\ln\Delta+\frac{3}{2},\
&& P_{2\Delta}=\Biggl(2\ln\Delta+\frac{3}{2}\Biggr)^2-4\zeta_2, \\ \nonumber
\Delta=\frac{\Delta\varepsilon}{\varepsilon}\ll 1,
&& \zeta_2=\frac{\pi^2}{6},
\ea
where $\rho_t$ is a so-called large logarithm,
$t$ is the kinematical invariant and $m_e,m_\mu$ --- masses of the leptons.

Besides we put the explicit form for the nonleading terms and present the result of lowest order RC calculation
to a power accuracy,
\be\label{acc}
1+{\cal O}\Biggl(\frac{\alpha}{\pi}\frac{m^2}{s}\rho_t\Biggr).
\ee
Our calculation of a second order contribution is done in the logarithmic approximation.
% \ba 1+{\cal O}\Biggl(\frac{1}{\rho_t}\biggr). \ea
We keep all the logarithmically enhanced terms including those containing logarithms of a mass ratio,
and omit the terms of order ${\cal O}(1)$.

Doing RC in the fourth order of PT we consider three separate gauge-invariant contributions.
We label them vertex contributions, decorated boxes and eikonal types. The last two
take into account the amplitudes with exchange between electron and muon enhanced
by additional one or two virtual
(or real soft) photons as well as virtual (real soft) pair.
Their contributions will be given elsewhere.

The first set of Feynman diagrams (FD) is that of the vertex type with RC to a second order (Fig.~\ref{fig2}).
It produces a contribution containing fourth power large logarithms along with the IR divergent terms.
Combining these with additional contributions coming from the emission and absorption of one and two soft photons
by either of the lepton lines results in the cancellation of fourth and third power of large logarithms as well as all
of IR divergent terms. The result is found to be in agreement with RG predictions.

%--------------- fig ---------------
\begin{figure}[!thb]
\begin{center}
\includegraphics[]{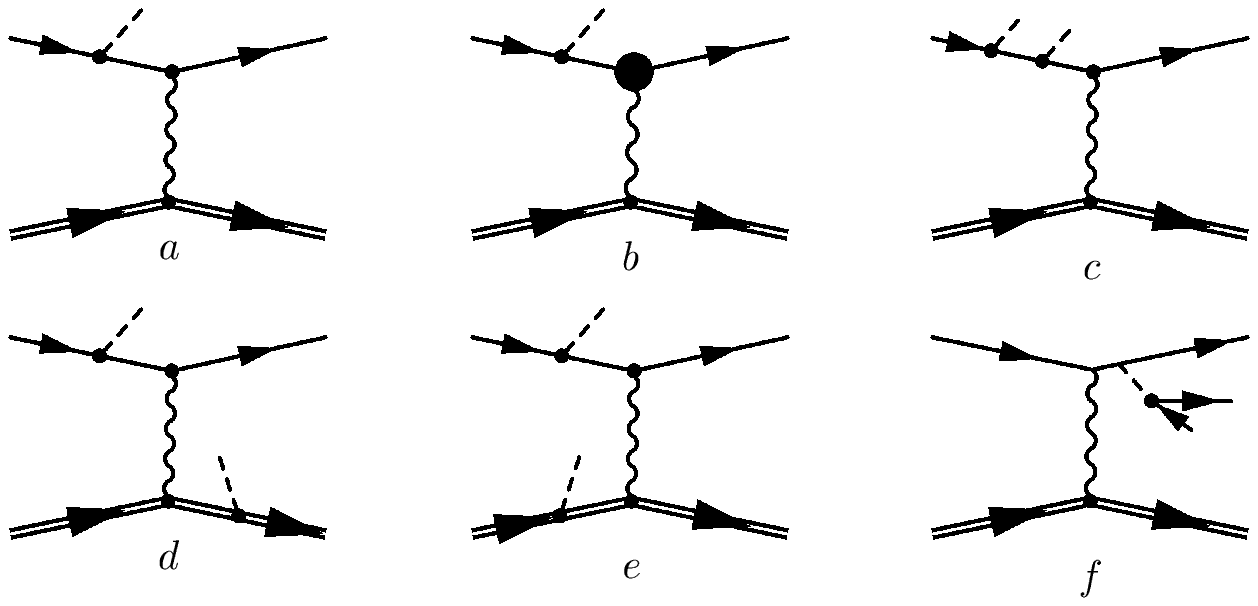}
\end{center}
\caption{ Some of the soft photon contributions: the diagram (a) corresponds to a first order RC;
in (b) a filled circle denotes vertex one-loop RC; (c,d,e) represent
an emission of two soft photons; (f) a soft pair production.}
\label{fig3}
\end{figure}
%--------------- fig ---------------

Our paper is organized as follows. After some introductory remarks we start with discussing a first order
contribution to the cross section of the process in Sec.~\ref{sec:2}.
In Sec.~\ref{sec:4} RC coming from the vertex diagrams to a $\alpha^4$ order of PT are
considered.
Section \ref{33} is devoted to the study of the vacuum polarization effects including
hadronic contribution to the vertex FD.
In Sec.~\ref{sec:3} we give the contribution due to the emission of one and two soft photons
as well as a soft pair for the cases of emitted (absorbed) leptons with equal
and different masses.
In conclusion we summarize and recapitulate the results obtained.

%--------------------------------------------------------------
\section{Born cross section and lowest order RC}
\label{sec:2}

Let's remind that here we are concerned with large-angle high-energy $e\mu$ scattering,
\ba
e^-(p_1)+\mu^-(p_2)&\to& e^-(p_1')+\mu^-(p_2'),\\ \nonumber
p_1^2=p_1^{'2}=m_e^2,&& p_2^2=p_2^{'2}=m_\mu^2,
\ea
with the kinematical invariants $s,t,u$ much larger than the lepton mass squared,
\[
s=(p_1+p_2)^2,\quad t=(p_1-p_1')^2=-\frac{s}{2}(1-c),
\quad u=(p_1-p_2')^2=-\frac{s}{2}(1+c),
\]
where $c=\cos(\widehat{\vecc{p}_1,\vecc{p}_1'})$
is a cosine of the scatter angle in cms (in what follows this reference frame is implied).
The differential cross section in the Born approximation has a form,
\ba \label{eqb}
\dd\sigma_0=\frac{1}{8s}B\dd\Gamma,\qquad
B=\sum|M_0|^2=8(4\pi\alpha)^2\frac{s^2+u^2}{t^2}, \\ \nonumber
\dd\Gamma=\frac{1}{(2\pi)^2}\frac{\dd^3p_1^{'}}{2\varepsilon_1}
\frac{\dd^3p_2^{'}}{2\varepsilon_2}
\delta^4(p_1+p_2-p_1^{'}-p_2^{'})=\frac{\dd\Omega_e}{8(2\pi)^2}.
\ea
And then we can write,
\[
\frac{\dd\sigma_0}{\dd\Omega_e}=\frac{\alpha^2}{2s}\frac{s^2+u^2}{t^2}
\Biggl\{1+{\cal O}\left(\frac{m_\mu^2}{s}\right)\Biggr\}.
\]

The lowest order RC comes from the emission of virtual (one-loop correction) and
real photons. The one-loop RC is classified out into the three distinct sets.
One of them is related with a vacuum polarization insertion into
the propagator of a photon exchanged between leptons. It could be taken into account as follows,
\ba
\left(\frac{\dd\sigma}{\dd\Omega_e}\right)_{vp}&=&\frac{\dd\sigma_0}
{\dd\Omega_e}\frac{1}{|1-\Pi(t)|^2},\\ \nonumber
\Pi(t)&=&\frac{\alpha}{3\pi}\left(l_t-\frac{5}{3}\right) +
\frac{\alpha}{3\pi}\left(L_t-\frac{5}{3}\right) + \delta_{had}(t) +
\frac{\alpha^2}{4\pi^2}(l_t+L_t) + \ldots\,, \\ \nonumber
\delta_{had}(t)&=&\frac{\alpha}{3\pi}\int\limits_{4m_\pi^2}
^{\infty}\frac{\dd M^2}{M^2}{\cal R}(M^2)\frac{t}{t-M^2} ,\qquad
l_t=\ln\frac{-t}{m_e^2}=\rho_t+L, \\ \nonumber L_t&=&\ln\frac{-t}{m_\mu^2}=\rho_t-L,
\qquad \qquad L=\ln\frac{m_\mu}{m_e}.
\ea
In the above the quantity $M^2$ denotes a square of hadron invariant mass in
a process $e\bar e\to h$,
\ba\label{eq:8}
{\cal R}(M^2)=\frac{\sigma_{e\bar e\to h}(M^2)}{\sigma_{e\bar e\to \mu\bar\mu}},
\ea
is the known ratio of the single-photon
annihilation cross sections with hadron and muon pairs produced.

Another set of one-loop RC contains vertex function (we remind that only the
Dirac form factor of the vertex function works within power accuracy quoted in Eq.~(\ref{acc})):
\ba\label{eq:7}
\left(\frac{\dd\sigma}{\dd\Omega_e}\right)_{v}&=&\frac{\dd\sigma_0}
{\dd\Omega_e}[V_e(l_t)V_\mu(L_t)]^2,\\ \nonumber
\ea
with lowest order Dirac form factors of leptons (see~\cite{BMR}):
\ba
V_e(l_t)&=&1+\frac{\alpha}{\pi}f^{(2)}_1(l_t)
+ \frac{\alpha^2}{\pi^2}f^{(4)}_1(l_t)+... ,
\qquad V_\mu(L_t)=V_e(l_t\to L_t),
% 1+\frac{\alpha}{\pi}f^{(2)}_1(L_t) + \frac{\alpha^2}{\pi^2}f^{(4)}_1(L_t),
\\ \nonumber
f^{(2)}_1(l_t)&=&l_\lambda(1-l_t) - 1 + \frac{3}{4}l_t
-\frac{1}{4}l_t^2 +\frac{1}{2}\zeta_2, \qquad l_\lambda=\ln\frac{m_e}{\lambda}.
\ea
Here $\lambda$ is a fictitious photon mass.
It's convenient to present $f^{(4)}_1(l_t)$ as a sum of two ingredients
\ba
\label{ssk}
f^{(4)}_1=f^{\gamma\gamma}+f^{vp},
\ea
one of which $f^{vp}$ contains QED vacuum polarization effects (it will be specified in section \ref{33})
and the second term,
\ba\label{fgg}
f^{\gamma\gamma}&=&\frac{1}{32}l_t^4-\frac{3}{16}l_t^3+\left(\frac{17}{32}
-\frac{1}{8}\zeta_2\right)l_t^2 +
\left(-\frac{21}{32}-\frac{3}{8}\zeta_2+\frac{3}{2}\zeta_3\right)l_t \\
\nonumber
&+&\frac{1}{2}l_\lambda^2(l_t-1)^2-l_\lambda(l_t-1)\left(-\frac{1}{4}l_t^2
+\frac{3}{4}l_t-1+\frac{1}{2}\zeta_2\right)+{\cal O}(1)\,,
\quad \zeta_3\approx 1.2020569\,.
\ea
A contribution due to the Pauli form factor is neglected for it is proportional
to a lepton mass squared. % of order $\frac{\alpha}{\pi}\frac{m_\mu^2}{t}L_t$.
Remaining one-loop RC is associated with the interference of the Born amplitude with those
containing two virtual photons exchanged between lepton lines.

A real photon emission could as usual be distinguished into a soft ($\omega<\Delta\varepsilon\ll\varepsilon$)
and a hard ($\omega>\Delta\varepsilon$) regions regarding photon energy.
For the quasireal case only a soft region is of relevance:
\ba
\frac{\dd\sigma_{soft}}{\dd\sigma_0}=\frac{-4\pi\alpha}{(2\pi)^3}
\int\frac{\dd^3\vecc{k}}{2\omega}R^2(k),\qquad
\omega=\sqrt{k^2+\lambda^2}<\Delta\varepsilon, \\ \nonumber
R(k)=Q_k^{p_1p_1'}+Q_k^{p_2p_2'}, \qquad Q_k^{pp'}=\frac{p'}{p'k}-\frac{p}{pk}.
\ea

Let's put some useful formul\ae{} for the description of a soft
photon emission. We imply the cms reference frame of initial particles, which
means that the values of 3-momenta of any particles (we consider $e\mu$ elastic
scattering) are equal.

First we give the expression for a single soft photon emission referred to
\ba
\label{eqs}
\delta^S_{11'}&=&-\frac{\alpha}{4\pi^2}\int\frac{\dd^3\vecc{k}}{\omega}
\left(\frac{p_1'}{p_1'k}-\frac{p_1}{p_1k}\right)^2\biggl|_{\omega<\Delta\varepsilon} \nonumber \\
&=&\frac{2\alpha}{\pi}\left[(l_t-1)\left(\ln\Delta+l_\lambda\right)
+\frac{1}{4}l_t^2-\zeta_2+\frac{1}{2}\Li\left(\frac{1+c}{2}\right)
\right],
\ea
with a dilogarithm function
\ba
\Li(z)=-\int\limits_0^z\frac{\dd x}{x}\ln(1-x). \nonumber
\ea
By a proper squaring of this formula one can easily derive the quantity $\delta^{SS}_{11'}$
(see formulae (\ref{ssq}) and (\ref{ssw})).

The IR free contributions to lowest order RC from two sets containing Dirac
form factors of the leptons along with the relevant contribution
coming from a soft photon emission could be cast into the following form,
which is in agreement with a structure function approach,
\ba
\frac{\dd\sigma^{(1)}}{\dd\sigma_0}&=&1+\frac{\alpha}{\pi}\bigl[\delta_{11^{'}}^S
+ 2f_1^{(2)}(l_t) + \delta_{22^{'}}^S + 2f_1^{(2)}(L_t)\bigr]= \\ \nonumber
&=&1 + \frac{\alpha}{\pi}
\left[2(\rho_t - 1)\left(2\ln\Delta+\frac{3}{2}\right) - 2\zeta_2 - 1 +
 2\Li\left(\frac{1+c}{2}\right)\right].
% B&=&8(4\pi\alpha)^2\frac{s^2+u^2}{t^2}.
\ea

After accounting for a soft photon contribution $\delta_{12^{'}}^S, \delta_{21^{'}}^S$
 as well as interference of the Born and box-type FD we obtain (details of lowest order box FD contribution
 can be found in \cite{Bhabha}):
\ba
\label{eqr}
\dd\sigma&=&\dd\sigma_0\Biggl\{
1 + \frac{\alpha}{\pi}\frac{1}{|1-\Pi(t)|^2}
\Biggl[\rho_t(4\ln\Delta+3) - 4\ln\Delta-4-2\zeta_2
+2\Li\left(\frac{1+c}{2}\right)\Biggr]+K\Biggr\}, \nonumber \\
K&=&\frac{\alpha}{\pi}\Biggl\{ L_{us}(4\ln\Delta+L_{st}+L_{ut})
+ 2\Li\left(\frac{1-c}{2}\right) + \frac{t^2}{s^2+u^2}
\Biggl[\frac{u}{t}L_{st} - \frac{s}{t}L_{ut} \\ \nonumber
&+&\frac{s-u}{2t}(6\zeta_2+L_{st}^2+L_{ut}^2)\Biggr]\Biggr\},
\ea
\[
L_{st}=\ln\frac{s}{-t},\quad L_{ut} = \ln\frac{u}{t},\quad L_{us} = \ln\frac{-u}{s}.
\]
The factor $K$ represents a sum of elastic Born-box amplitudes and corresponding inelastic contributions.
The expression for the cross section given above is in agreement with predictions expected from RG considerations.

The LLA expression for the EMS cross section can be brought to the form
of a Drell-Yan-like process~\cite{FK} written in terms of structure functions:
\ba \label{DY}
\dd\sigma=\frac{\dd\sigma_0}{|1-\Pi(t)|^2}
\left[{\cal D}_{\Delta}\left(\frac{\alpha(t)}{2\pi}l_t\right)\right]^2
\left[{\cal D}_{\Delta}\left(\frac{\alpha(t)}{2\pi}L_t\right)\right]^2,
\ea
with the non-singlet structure functions ${\cal D}_\Delta$,
\ba
{\cal D}_\Delta(z)=1+zP_{1\Delta}+\frac{1}{2}z^2P_{2\Delta}+\ldots+\frac{1}{n!}z^nP_{n\Delta}+\ldots
\ea
Here $P_{n\Delta}$ is the n$^{th}$ iteration $\Delta$-part of the kernel of
evolution equations:
\ba
P_n(y)&=&\lim_{\Delta\to 0}[\,P_{n\Delta}\delta(1-y)+\Theta(1-y-\Delta)P_{n\theta}\,]
=\int\limits_y^1\frac{\dd x}{x}P_1(x)P_{n-1}\left(\frac{y}{x}\right), \qquad n\geq 2,
\nonumber \\
P_{1\theta}&=&\frac{1+y^2}{1-y},\quad P_{2\theta}=\frac{1+y^2}{1-y}\left[
\ln\frac{(1-y)^2}{y}+\frac{3}{2}\right]+\frac{1}{2}(1+y)\ln y - (1-y).
\ea
Explicit expressions for $P_{1\Delta}, P_{2\Delta}$ are given above in (\ref{eq1}).
Parameter $\Delta$ $(\Delta\ll 1)$ can be interpreted as energy fraction carried by
soft real photons and pairs escaping detectors. $\alpha(t)$ is the running QED coupling constant
$
\alpha(t)=\alpha/(1-\frac{\alpha}{3\pi}t).
$

%--------------------------------------------------------------
\section{ Second order RC }
\label{sec:3}

These can be represented as a sum of several sets each of which depending
on a choice of gauge with respect to virtual as well as real photons.
We consider FD describing elastic scattering
with vacuum polarization effects included and furthermore with
account for a soft pair production.
They are related with the one photon exchange FD both for
elastic and quasi-elastic processes and could be specified by
the emission of two more (either virtual or real) photons
out of the same lepton lines.

A keystone to this classification is a soft photon radiator cross section.
In case of only one soft photon emitted it gets the form,
\ba
&&\dd\sigma_{soft}=\frac{1}{8s}\frac{\dd\Omega_e}{8(2\pi)^5}\,
(\delta\sum|M|^2)_{soft}\,\frac{\dd^3\vecc{k}}{2\omega}\biggr|_{\frac{2\omega}{\sqrt{s}}<\Delta}, \\ \nonumber
&&(\delta\sum|M|^2)_{soft}=2\Re e\sum M_0^*M^{(1)}(-4\pi\alpha)R^2(k).
\ea
For the two soft photon emission (for instance by electron block) we have ,
\ba
\label{ssq}
\dd\sigma_{ss}=\dd\sigma_0\frac{1}{2!}\left(-4\pi\alpha\right)^2
\frac{\dd^3\vecc{k}_1}{2\omega_1(2\pi)^3}
\frac{\dd^3\vecc{k}_2}{2\omega_2(2\pi)^3}
(Q_{k_1}^{p_1p_1^{'}})^2(Q_{k_2}^{p_2p_2^{'}})^2\biggr|_{\frac{2\omega_1}{\sqrt{s}}+\frac{2\omega_2}{\sqrt{s}}<\Delta}.
\ea

For the case of emission of two soft photons provided that their total
energy does not exceed $\Delta\varepsilon\ll\varepsilon$ we have,
\ba
\label{ssw}
\Biggl[\int\frac{\dd^3\vecc{k}_1}{\omega_1}\frac{p_ip_j}{p_ik_1\cdot p_jk_1}
\cdot\int\frac{\dd^3\vecc{k}_2}{\omega_2}\frac{p_lp_m}{p_lk_2\cdot p_mk_2}
\Biggr]\Bigg|_{\omega_1+\omega_2<\Delta\varepsilon}=
(a_1\ln\Delta+b_1)(a_2\ln\Delta+b_2)-a_1a_2\zeta_2.
\ea
In the above the following integrals are defined,
\ba
\left[\int\frac{\dd^3\vecc{k}_1}{\omega_1}\frac{p_ip_j}{p_ik_1\cdot p_jk_1}
\right]\Bigg|_{\omega_1<\Delta\varepsilon}=a_1\ln\Delta + b_1, \nonumber \\
\left[\int\frac{\dd^3\vecc{k}_2}{\omega_2}\frac{p_lp_m}{p_lk_2\cdot p_mk_2}
\right]\Bigg|_{\omega_2<\Delta\varepsilon}=a_2\ln\Delta + b_2. \nonumber
\ea

The general structure of all the above contributions to the differential
cross section reveals a presence of large logarithms up to a fourth power.
However in overall sum one observes only their second powers.
Such a cancellation is characteristic of each gauge invariant set of corrections.
%--------------------------------------------------------------
\subsection{Vertex graphs}
\label{sec:4}

Three gauge invariant groups of FD containing one photon exchange contribute
\ba
\label{eq12}
\frac{\dd\sigma^v}{\dd\sigma_0}=\frac{\alpha^2}{\pi^2}[a_1+\tilde{a}_1+a_2].
\ea
The quantity $\tilde a_1$ is related with the emission of two (virtual and real)
photons out of a muon line: $\tilde a_1=a_1(l_t\to L_t)$.

Using results given in Eq.~(\ref{fgg}) for the electron Dirac form factor up to
a fourth order of PT
\footnote{We exclude a contribution due to the vacuum polarization, which will be taken into
account in what follows.}, the following IR singularities free contributions to the matrix
element squared from one photon exchange amplitudes could be constructed,
\ba\label{eq:14}
a_1&=&(f_1^{(2)})^2 + 2f^{\gamma\gamma} + 2f_1^{(2)}\delta^S_{11'}
+ \delta^{SS}_{11'},\\ \nonumber
a_2&=&4f_1^{(2)}\tilde{f}_1^{(2)} + 2[f_1^{(2)}\delta^S_{22'}+
\tilde{f}_1^{(2)}\delta^S_{11'}] + \delta^S_{11'}\delta^S_{22'},
\ea
where $\tilde{f}^{(2)}_1$ corresponds to a muon form factor identical to an
electron one with electron mass replaced by that of a muon.
The quantities $\delta^S_{ij}$ and $\delta^{SS}_{ij}$ correspond to
the emission of one and two soft real photons (their energy are restricted by
condition $\Delta\omega_1+\Delta\omega_2<\epsilon$) off fermion lines $i,j$.
The corresponding expression is given in Eq.~(\ref{eqs}).
One should take note of the factor $1/2!$ in front of the latter quantities
that is due to the identity of the soft photons emitted.

The relevant contribution to the differential cross section in logarithmic
approximation then appears to be,
\ba
\label{eq13}
a_1 + \tilde a_1&=& \rho_t^2P_{2\Delta} + \rho_t\left[-\frac{45}{8}+Y + 2\zeta_2 + 6\zeta_3\right]
 + {\cal O}(1), \\ \nonumber
a_2&=&\rho_t^2P_{2\Delta} + \rho_t\left[-6 + Y + 5\zeta_2 \right] + {\cal O}(1),\\
\nonumber
 Y&=&2P_{1\Delta}\Li\left(\frac{1+c}{2}\right)
-(4\zeta_2+14)\ln\Delta - 8\ln^2\Delta,
\ea
quantities $P_{1,2\Delta}$ are defined in (\ref{eq1}).
This result is in agreement with RG form of large-angle cross section.

%--------------------------------------------------------------
\subsection{Hadronic vacuum polarization}
\label{33}

We study vacuum polarization effects occurring while considering
vertex FD (see Fig.~\ref{fig2}b,c). To this end,
the known expression for hadronic vacuum contribution to the photon Green
function is used by doing the following substitution,
\ba
\frac{1}{k^2}\longrightarrow \frac{\alpha}{3\pi}\int\limits_{4m_\pi^2}^{\infty}
\frac{\dd M^2}{M^2}\frac{ {\cal R}(M^2)}{k^2-M^2},
\ea
where $k$ is the four-momentum of the virtual photon, $M^2$ is a hadron invariant mass squared
and the ratio ${\cal R}(M^2)$ is given in Eq.~(\ref{eq:8}).

In the next order of PT we must consider the three gauge invariant classes of FD
for elastic and quasi-elastic processes with a soft photon and a soft pion pair production.
At first the vertex class is examined. They could be written as
\ba
\frac{\dd\sigma}{\dd\sigma_0}=1+(\delta_s+\delta_v)^{hadr},&& \delta_v^{hadr}=\frac{\alpha^2}{6\pi^2}
[F(m_e^2,t)+F(m_\mu^2,t)], \\ \nonumber
F(m^2,t)=\int\limits_{4m_\pi^2}^{\infty}\frac{\dd M^2}{M^2}\!\!\!&&\!\!\!
{\cal R}(M^2)\,F_1(t,m^2,M^2),
\ea
where $F_1$ is the vertex (with hadronic vacuum polarization of virtual photon) contribution
to the Dirac form factor of a lepton having mass $m$. The
contribution of the Pauli form factor $F_2$ is suppressed by a factor of $|m^2/t|$.
$\delta_s^{hadr}$ corresponds to soft hadron's emission of soft pion pairs.

A standard calculation with the regularization at $t=0$ leads to (for details see App. A):
\ba
F_1(t,m^2,M^2)=2\int\limits_0^1\dd x\int\limits_0^1y\dd y \biggl[\ln\frac{d_0}{d}+\frac{a}{d}-\frac{a_0}{d_0}\biggr],
\ea
with
\ba
\label{quan}
a=a_0+t[1-y+x(1-x)y^2],&& d=d_0-y^2x(1-x)t, \\ \nonumber
a_0=-m^2(2-y^2),&& d_0=y^2m^2+(1-y)M^2.
\ea
It can be seen that the condition $F_1|_{t=0}=0$ is satisfied.
Here we put two limiting cases for $F_1$. In the case of large  hadron invariant mass squared,
as compared to $-t$, it's found to be:
\be
F_1(t,m^2,M^2)=\frac{t}{M^2}\biggl[\frac{2}{3}\ln\frac{M^2}{-t}+\frac{11}{9}\biggr],
\qquad M^2\gg -t,
\ee
and for the case of small invariant mass squared:
\be
F_1(t,m^2,M^2)=-\ln^2\frac{-t}{m^2}-2\ln\frac{M^2}{m^2}\ln\frac{-t}{m^2}-5\ln\frac{-t}{m^2}+\frac{\pi^2}{3}-\frac{1}{2},
 \qquad -t\gg M^2\gg m_\mu^2.
\ee
Taking into account the emission of soft pairs (see App. B) we have for hadronic contribution
radiative correction:
\ba
\label{hadr}
(\delta_v+\delta_s)^{hadr}=\frac{\alpha^2}{6\pi^2}\int\limits_{4m^2_{\pi^2}}^{-t}
\frac{\dd M^2}{M^2}R(M^2)\biggl[-\ln\frac{-t}{M^2}\biggl[8\ln\frac{M^2}{m_em_\mu}-2\ln\Delta+10\biggr] \\ \nonumber
-6\ln^2\frac{M^2}{m_em_\mu}-10\ln\frac{M^2}{m_em_\mu}-6\ln^2\frac{m_\mu}{m_e}+\frac{2}{3}\pi^2-1\biggr].
\ea

\subsection{Leptonic vacuum polarization, soft lepton pairs.}
\label{vp}

Next we do a VP-type contribution to the lepton vertex function.
Obviously, there are two possibilities for a VP blob to be inserted into
the lepton vertex function.
Then the contribution to elastic cross section could be written as,
\ba\label{fr}
\left(\frac{\dd\sigma^{vp}}{\dd\sigma_0}\right)_e
=2\frac{\alpha^2}{\pi^2}\Biggl[Z_1(m_e,m_e) + Z_2(m_e,m_\mu)\Biggr],
\ea
where
\ba
Z_1(m_e,m_e)=-\frac{1}{36}\rho_t^3 +\frac{1}{12}\left(\frac{19}{6}-L\right)\rho_t^2
- \frac{1}{36}\left(6 \zeta_2 + \frac{265}{6}+3 L^2-19 L\right)\rho_t\equiv f^{vp},\nonumber
\ea
is a contribution of the electron blob inserted into the electron vertex function  (for
definition $f^{vp}$ see (\ref{ssk})) and
\ba
Z_2(m_e,m_\mu)=-\frac{1}{36}\rho_t^3 + \frac{1}{12}\left(\frac{19}{6}+L\right)\rho_t^2
-\frac{1}{36}\left(6 \zeta_2 + \frac{265}{6}+3 L^2+63 L\right)\rho_t,      \nonumber
\ea
is a muon blob contribution to the electron vertex respectively.

The similar expression holds for muon vertex function (electron blob contribution for muon vertex):
\ba
Z_3(m_\mu,m_e)=-\frac{1}{36}\rho_t^3 + \frac{1}{12}\left(\frac{19}{6}-L\right)\rho_t^2
-\frac{1}{36}\left(6 \zeta_2 + \frac{265}{6}+3 L^2-25 L\right)\rho_t,  \nonumber
\ea

Now turn to inelastic process of a soft lepton-antilepton (of mass $\mu$
obeying $2\mu\ll\Delta\varepsilon\ll\varepsilon$) pair production.
For the differential cross section we obtain:
\ba
\label{fra}
\frac{\dd\sigma^{sp}}{\dd\sigma_0}=\frac{\alpha^2}{6\pi^2}
\Biggl[\frac{1}{3}\mathbf{L}^3 + \mathbf{L}^2\left(2\ln\Delta-\frac{5}{3}\right)
+\mathbf{L}\Biggl(4\ln^2\Delta-\frac{20}{3}\ln\Delta+\frac{56}{9}
-4\zeta_2+2\Li\left(\frac{1+c}{2}\right)\Biggr)\Biggr],
\ea
with $\mathbf{L}$ defined as
$
\mathbf{L}=\ln(-t/\mu^2).
$
We assume a muon or an electron to be a scattered lepton,
consequently the quantity $\mu$ stands for the corresponding mass.

The sum of contributions~(\ref{fr}) and (\ref{fra}) doesn't contain cubic
powers of large logarithms and for the ``electron line corrections''
is found to be (see~\cite{AAE}):
\ba\label{sa}
\left(\frac{\dd\sigma_{vp}^{sv}}{\dd\sigma_0}\right)_e
&=&\left(\frac{\alpha}{\pi}\right)^2
 \Biggl\{\biggl(\frac{2}{3}\ln\Delta + \frac{1}{2}\biggr) \rho^2_t
\nonumber \\
&+&2\rho_t\Biggl[-\frac{17}{12}-\frac{11}{9}L+\frac{2}{3}\ln^2\Delta - \frac{10}{9}\ln\Delta
-\zeta_2+\frac{1}{3}\Li\left(\frac{1+c}{2}\right)\Biggr]\Biggr\}.
\ea
In case of a muon it appears to be,
\ba \label{sb}
\left(\frac{\dd\sigma_{vp}^{sv}}{\dd\sigma_0}\right)_\mu
&=&\left(\frac{\alpha}{\pi}\right)^2
 \Biggl\{\biggl(\frac{2}{3}\ln\Delta + \frac{1}{2}\biggr) \rho^2_t
 \nonumber \\
&+&2\rho_t\Biggl[-\frac{17}{12}+\frac{11}{6}L+\frac{2}{3}\ln^2\Delta-\frac{10}{9}
\ln\Delta - \zeta_2+\frac{1}{3}\Li\left(\frac{1+c}{2}\right)\Biggr]\Biggr\}.
\ea
It's seen that the leading terms are in agreement with RG predictions.

\section{Summary}
\label{con}
In this paper we evaluate the Born cross section and the first order RC
to it of a process of $e\mu$ scattering in the quasi-elastic kinematical situation.
The relevant formulae are given in LLA in (\ref{eq1},\ref{eq1a}) and with power accuracy in (\ref{eqr}).

Among second order contributions we consider gauge-invariant contributions from FD with radiative corrections to
vertex function of either leptons. Here we also include soft photon and pairs emission with energies
less than $\Delta\epsilon$.

In ~LLA the results are in agreement with RG.

The explicit results for virtual and soft real photon emission are given in formulae (\ref{eq12},\ref{eq13}).
For emission of virtual and soft real lepton pairs the relevant formulae are given in (\ref{fra},\ref{sa}).

Section \ref{33} in this paper is devoted to the determination of the contributions coming
from the hadronic vacuum polarization, where RC is expressed in terms of explicit integral from the
experimentally measured quantity $R(M^2)$. Also we consider a soft pion pair production (see App.B).
We calculate the hadronic vacuum polarization
contribution to the vertex functions of electron or muon explicitly.
The relevant formulae for RC are given in (\ref{hadr}).

In the forthcoming papers we will consider contributions coming
from the last gauge-invariant types of contributions, which are eikonal
and decorated types of box FD.

\section*{Appendix~A. Hadronic vacuum polarization, details}

Let's consider here details of vertex hadron's function calculation. For the vertex function we can write down:
\ba
V_\mu=\Gamma_1\gamma_\mu+\Gamma_2(\hat{q}\gamma_\mu-\gamma_\mu\hat{q}),
\ea
where $q=p_2-p_2^{'}$ and $\Gamma_{1,2}$ are Dirac and Pauli form-factor.
Let's write down vertex function as follows,
\ba
V_\mu=\gamma_\mu[\Gamma_1+4m\Gamma_2]-2(p_2+p_2^{'})_\mu\Gamma_2=\gamma_\mu A+(p_2+p_2^{'})_\mu B,
\ea
where
\ba
A&=&\int y\dd y\int\dd x
\biggl[\frac{2y2p_2p_2^{'}}{-d}+\frac{4y^2}{d}(m^2+p_2p_2^{'})-2y\ln\frac{d}{m^2} \\ \nonumber
&+&\frac{-2m^2y^3x^2}{d}+\frac{2y^3x(1-x)}{d}(-2p_2p_2^{'})\biggr], \\
B&=&\int y\dd y\int \dd x\biggl[\frac{y^2}{d}(-2m)+\frac{2y^3x^2}{d}2m+\frac{2y^3x(1-x)}{d}2m\biggr].
\nonumber
\ea
Quantities $d,d_0$ are defined in Eq.~(\ref{quan}). After a regularization at $t=0$ we have:
\ba
F_1(t,m^2,M^2)=\Gamma_1-\Gamma_1|_{t=0}=2\int\limits_0^1\dd x\int\limits_0^1y\dd y \biggl[\ln\frac{d_0}{d}
+\frac{a}{d}-\frac{a_0}{d_0}\biggr].
\ea
The contribution of the Pauli form factor $\Gamma_2$ is proportional to $B$ and
therefore is suppressed by a factor of $|m^2/t|$.

\section*{Appendix~B. Soft pion pair production}
\label{hadron}
The general expression for a soft pion pair production,
\ba
\left|\frac{M}{M_0}\right|^2\dd\Gamma_{\pm}&=&\left(\frac{4\pi\alpha}{q^2}\right)^2
\dd^4q\int\frac{\dd^3 q_+}{2 \varepsilon_+}\int\frac{\dd^3 q_-}{2 \varepsilon_-}
(2\pi)^{-6}\delta^4(q_++q_--q)          \nonumber \\
&\times&(q_+-q_-)_\mu(q_+-q_-)_\nu J_\mu J_\nu,\quad
J_\mu=(Q_q^{p_1p_1'})_\mu, \nonumber
\ea
with
\[
m_\mu\ll\sqrt{q^2}\ll\Delta\varepsilon\ll\varepsilon,\quad q_0^2\gg q^2.
\]
Here $q_\pm, \epsilon_\pm$ is four-momentum and energy of $\pi^\pm$, $q$ and $q_0$ is 4-momenta
and energy of soft pair.

Rewriting
\ba
\int\dd^4q&=&\frac{4\pi}{2}\int\dd q^2\frac{\dd\Omega_q}{4\pi}
\int\limits_{\sqrt{q^2}}^{\Delta\varepsilon}\dd q_0\sqrt{q_0^2-q^2}, \nonumber \\
\int\frac{\dd^3 q_+}{2 \varepsilon_+}\!\!\!\!\!\!&&\!\!\!\!\!\!
\int\frac{\dd^3 q_-}{2 \varepsilon_-}
(2\pi)^{-6}\delta^4(q_++q_--q)(q_+-q_-)_\mu \nonumber \\
\!\!\!\!\!&\times&\!\!\!\!\!(q_+-q_-)_\nu =\frac{1}{3}
\left(g_{\mu\nu}-\frac{q_\mu q_\nu}{q^2}\right)
2^{-7}\pi^{-5}(4m^2-q^2)\sqrt{1-\frac{4m^2}{q^2}}, \nonumber
\ea
one gets,
\ba
\left|\frac{M}{M_0}\right|^2\dd\Gamma_{\pm}=-\frac{\alpha^2}{4\pi^3}\frac{(q^2-4m^2)^{3/2}}
{(q^2)^3}\int\frac{\dd\Omega_q}{4\pi}\int\dd q_0\sqrt{q_0^2-q^2}J^2, \nonumber
\ea
where
\[
J^2=J_\mu J_\nu\left(g_{\mu\nu}-\frac{q_\mu q_\nu}{q^2}\right).
\]
Separate contributions,
\ba
\int\frac{\dd\Omega_q}{4\pi}\frac{m^2}{(p_1q)^2}&=&{\cal O}\left(\Delta^2
\frac{m^2}{q^2}\right),\nonumber \\
\int\frac{\dd\Omega_q}{4\pi}\frac{p_1p_2}{p_1qp_2q}&=&
\int\frac{\dd\Omega_q}{4\pi}\frac{p'_1p'_2}{p'_1qp'_2q}=
\frac{1}{2}\ln^2\left(\frac{2\Delta\varepsilon}{\sqrt{q^2}}\right)-\ln 2,\nonumber
\ea
Master integral,
\ba
\int\limits_{\sqrt{q^2}}^{\Delta\varepsilon}\dd q_0\sqrt{q_0^2-q^2}
\int\frac{\dd\Omega_q}{4\pi}\frac{p_1p'_1}{p_1qp'_1q}=
\ln^2\left(\frac{2\Delta\varepsilon}{\sqrt{q^2}}\right)
+\ln\left(\frac{2\Delta\varepsilon}{\sqrt{q^2}}\right)\ln\left(\frac{1-c}{2}\right)
-\zeta_2,\nonumber
\ea
Soft pion pair production contribution to the invariant mass distribution
(both from electron and muon blocks emission) has the form:
\ba
\frac{M^2}{\sigma_0}\frac{\dd\sigma}{\dd M^2}=\frac{\alpha^2}{3\pi^2}
\left[\ln^2\frac{-t}{M^2}+\ln\frac{-t}{M^2}\ln\frac{\Delta\varepsilon}{\varepsilon}
+{\cal O}(1)\right].\nonumber
\ea

%%%%%%%%%%%%%%%%%%%%%%%%%%%%%%%%%%%%%%%%%%%%%%%%%%%%%%%%%%%%%%%%%%%%%%%%%%

\end{document}